# Thermal Magnetoelectrics in all Inorganic Quasi-Two-Dimensional Halide Perovskites

Tong Zhu[1]†, Xuezeng Lu[2]†, Takuya Aoyama[3], Koji Fujita[4], Yusuke Nambu[5,6,7], Takashi Saito[8], Hiroshi Takatsu[1], Tatsushi Kawasaki[4], Takumi Terauchi[4], Shunsuke Kurosawa[5,9,10], Akihiro Yamaji[5,9], Hao-Bo Li[11,12], Cedric Tassel[1], Kenya Ohgushi[3], James M. Rondinelli[2]*, Hiroshi Kageyama[1]*

[1]Department of Energy and Hydrocarbon Chemistry, Graduate School of Engineering, Kyoto University, Nishikyo-ku, Kyoto, 615-8510, Japan.

[2]Department of Materials Science and Engineering, Northwestern University, Evanston, Illinois, 60208, USA.

[3]Department of Physics, Graduate School of Science, Tohoku University, Sendai, 980-8578, Japan.

[4]Department of Material Chemistry, Kyoto University, Nishikyo-ku, Kyoto, 615-8510, Japan.

[5]Institute for Materials Research, Tohoku University, Sendai, 980-8577, Japan.

[6]Organization for Advanced Studies, Tohoku University, Sendai, 980-8577, Japan.

[7]FOREST, Japan Science and Technology Agency, Kawaguchi, Saitama, 332-0012, Japan.

[8]Institute of Materials Structure Science, High Energy Accelerator Research Organization (KEK), Tokai, Ibaraki, 319-1106, Japan.

[9]New Industry Creation Hatchery Center (NICHe), Tohoku University, Sendai, 980-8578, Japan.

[10]Institute of Laser Engineering, Osaka University, Suita, Osaka, 990-8560, Japan.

[11]SANKEN, Osaka University, Ibaraki, Osaka, 567-0047, Japan.

[12]Spintronics Research Network Division, Institute for Open and Transdisciplinary Research Initiatives, Osaka University, Suita, Osaka, 565-0871, Japan.

*Corresponding author. Email: kage@scl.kyoto-u.ac.jp (H.K.); jrondinelli@northwestern.edu (J.M.R.)

†These authors contributed equally to this work

## Introductory Paragraph

From lithium-ion batteries[1] to high-temperature superconductors[2], oxide materials have been widely used in electronic devices. However, demands of future technologies require materials beyond oxides, as anion chemistries distinct from oxygen can expand the palette of mechanisms and phenomena, to achieve superior functionalities. Examples include nitride-based wide bandgap semiconductors and halide perovskite solar cells, with MAPbBr$_3$ being a representation revolutionizing photovoltaics research[3]. Here, we demonstrate magnetoelectric behaviour in quasi-two-dimensional halides (K,Rb)$_3$Mn$_2$Cl$_7$ through simultaneous thermal control of electric and magnetic polarizations by exploiting a polar-to-antipolar displacive transition.  Additionally, our calculations indicate a possible polarization switching path including a strong magnetoelectric coupling, indicating halides can be excellent platforms to design future multiferroic and ferroelectric devices. We expect our findings to broaden the exploration of multiferroics to non-oxide materials and open access to novel mechanisms, beyond conventional electric/magnetic control, for coupling ferroic orders.

**Main Texts**

Simultaneous thermal control of the two coupled ferroic orders in magnetoelectric materials offers possibilities for new technological inventions such as multiple-state information storage devices[4,5]. In addition, the cross-control of electric and magnetic polarizations, using an electric or magnetic field, makes it possible to design non-volatile, energetically efficient magnetic storage devices[6,7]. However, the development of magnetoelectric materials is hampered by intrinsic chemical incompatibility [8]. Specifically, non-centrosymmetric, polar crystal structures needed for electric polarization are typically stabilized by second-order Jahn-Teller distortions[9] of closed-shell cations with either $d^0$ or $ns^2$ electronic configurations[8,10,11]. These electronic structures are incompatible with magnetic properties which require unpaired electrons. Multiple approaches have been utilized to circumvent the contra-indication. In Type-I multiferroics, polar distortions and magnetism are introduced from different cations, typically in complex oxides, e.g., $BiFeO_3$[12]. Because polar distortions and magnetism have different origins, the magnetoelectric coupling between two ferroic orders is usually weak. Alternative single-phase multiferroics such as rare-earth manganates rely on polar noncollinear magnetic spin structures[13,14], which are categorized as Type-II multiferroics, yet the polarization is usually small.

Recently, hybrid-improper ferroelectricity (HIF) has emerged as an alternative and promising mechanism to realize multiferroics with strong magnetoelectric coupling[15,16]. Under this mechanism, typically in quasi-2D perovskite oxides such as $A_3B_2O_7$ Ruddlesden-Popper (RP) phases, two non-polar octahedral rotations couple together and give rise to electric polarizations[17,18]. This mechanism has two advantages: First, polar structures arise from size-mismatch between A- and B-site cations, and there is no special electronic structure requirement. Therefore, it is tactically simpler to unify electric polarizations with long-range magnetic order.

Second, both electric and magnetic polarizations are coupled to the same structural distortions, from which a strong coupling between the two ferroic orders is anticipated; this effect leads to the corollary that the polarization is usually larger than Type-II multiferroics. Despite these merits, only three candidates, all of which are oxides, have been identified, including $Ca_3Mn_2O_7$[15,19], $[Ca_{0.69}Sr_{0.46}Tb_{1.85}Fe_2O_7]_{0.85}[Ca_3Ti_2O_7]_{0.15}$[20] and $MnSrTa_2O_7$[21].

Here we prepared and examined a series of the first inorganic halide HIF magnetoelectric materials $(K,Rb)_3Mn_2Cl_7$ with quasi-2D-perovskite structures. Employing halides rather than oxides offers a better chance to combine two ferroic orders, because the large total negative charge from oxide anions requires high valent B-site cations (typically 4+ or 5+) and this limits the number of paramagnetic (PM) cations one can choose. In contrast, divalent magnetic cations can be easily incorporated into the B-site of halide perovskites due to the lower charge of the halide anions. The presence of a large number of divalent magnetic cations suitable for octahedral geometry, together with the wide variety of nonmagnetic monovalent cations (such as alkali metals), enable a rational design of a large family of new HIF multiferroics. We then show simultaneous thermal manipulation of the electric and magnetic polarizations in these materials. This new phenomenon is triggered by a polar-to-antipolar structural transition, which has never been observed in HIF oxide ferroelectrics/multiferroics, and provides a thermally-driven route to circumvent the challenge of lowering the electric coercive field energy demanded by future microelectronic devices[22].

Polycrystalline $K_3Mn_2Cl_7$ was synthesized via a ceramic synthesis route from KCl and $MnCl_2$, with a modified condition from a previous report[23]. Refinements against low-temperature synchrotron X-ray diffraction (SXRD) and neutron powder diffraction (NPD) data (Supplementary Information Fig. S1) reveal it adopts a polar, $a^-a^-c^+/a^-a^-c^+$ distorted, $n = 2$ RP structure, in space group $A2_1am$ (Fig. 1A). This low-temperature polar structure is unambiguously demonstrated by pyroelectric and second-harmonic generation data (Fig. 1C, Supplementary Information Fig. S30). It arises from a HIF mechanism by combining $a^-a^-$ out-of-phase rotations ($X_3^-$ irreducible representation, irrep), $c^+$ in-phase rotation ($X_2^+$ irrep), and a polar $\Gamma_5^-$ distortion through a trilinear coupling interaction $Q_{X_3^-}Q_{X_2^+}Q_{\Gamma_5^-}$ (Supplementary Information Figs. S2-S4). On warming, $K_3Mn_2Cl_7$ undergoes a first-order transition between $T_{C1}$ ~ 155 K and $T_{C2}$ ~ 180 K to adopt a hybrid-improper antipolar $P4_2/mnm$ structure with $a^-b^0c^0/b^0a^-c^0$ distortion ($X_3^-$ irrep), which stabilizes an antipolar distortion mode ($M_2^+$) leading to local polarization in each perovskite sheet but cancel within the layer (Supplementary Information Figs. S5-S8). Further warming to $T_A$ ~ 410 K leads to a transition to the undistorted $I4/mmm$ structure (Figs. 1A,1B, Supplementary Information Figs. S9, S10).

Zero field-cooled (ZFC) and field-cooled (FC) magnetization data (Fig. 2A) split at 64 K, while NPD data collected below 64 K (Fig. 2B) reveals a series of sharp magnetic reflections, indicating a long-range magnetic order below 64 K ($T_N$). Strong diffuse magnetic features in the NPD suggesting short-range spin correlations are observed between 64 K and 100 K, consistent with a broad maximum in magnetic susceptibility around 100 K ($T_{max}$). Fitting of the 4 K NPD data (Fig. 2C, Supplementary Information Fig. S11) reveals a G-type antiferromagnetic order in magnetic space group $A2_1'am'$, with refined moments (4.36(2) $\mu_B$) aligning along the crystallographic $c$-axis (Fig. 2E). This magnetic symmetry permits weak ferromagnetic (wFM) canting along the $b$-axis via the Dzyaloshinskii-Moriya ($DM$) interaction[24] (Fig. 2E), as evidenced by non-linear hysteretic magnetization-field isotherms (Fig. 2D) and density functional theory (DFT) calculations (Supplementary Information Figs. S17, S18). Fitting the temperature-dependence of the ordered moments using a power law $M = M_0(1 - \frac{T}{T_N})^\beta$ yields a Néel temperature of $T_N$ = 64.3(1) K (Fig. 2F).

The coupling between the electric polarization and the wFM order in $K_3Mn_2Cl_7$ is established by examining the crystal and magnetic symmetries (Fig. 3A). The electric polarization ($P$) is directly coupled to the $X_3^-$ out-of-phase tilt and $X_2^+$ in-phase rotation. The ferromagnetic order (m$\Gamma_5^-$ irrep) is established via a trilinear coupling mechanism with the $X_3^-$ tilt and $mX_1^-$ G-type antiferromagnetic order. Consequently, two ferroic orders are directly coupled via the $X_3^-$ tilt, and can be simultaneously manipulated by applying an external stimulus which selectively modifies this distortion. A close inspection of the NPD data reveals a clear anomaly (~ 2% drop) of the polar $\Gamma_5^-$ distortion at $T_N$ ~ 64 K (Fig. 3B), indicating the electric polarization responds to the onset of 3D magnetic order, hence a coupling effect. Another noticeable observation is the negative thermal expansion (NTE) of the polar $a$-axis below $T_N$, indicative of strong magnetoelastic coupling and a situation not reported in previous HIF multiferroic materials.

The coupling scheme established above is analogous to $Ca_3Mn_2O_7$[15], but crucially differs in switching pathways. In $Ca_3Mn_2O_7$, the lowest energy pathway to switch the polarization with the change of $X_3^-$ rotation cannot be unambiguously established and simultaneous manipulation of polarization and magnetism has not been observed. In contrast, our calculations show the lowest energy path for $K_3Mn_2Cl_7$ involves switching the $X_3^-$ tilt, which transforms via an antipolar intermediate $P4_2/mnm$ state (Fig. 3C). This pathway is lower in energy than transition paths through the 90° twin boundary and $Pnma$ phase that have been reported in HIF RP oxides[25,26] by 4.2 meV/f.u. and 3.6 meV/f.u., respectively. To the best of our knowledge, this calculated pathway has never been unambiguously demonstrated in oxide HIF ferroelectrics/multiferroics and it is likely to also switch the wFM in $K_3Mn_2Cl_7$ (Fig. 3A, Supplementary Information Figs. S26-S28). This unusual polarization-switching behaviour may be attributed to the anion chemistry and suggests that halide HIF multiferroic materials can be good platforms to simultaneous switch electric and magnetic polarisations.

The suggested intermediate $P4_2/mnm$ phase in the polarization switching path identified from our DFT calculations is in excellent agreements with the observed thermal phase transitions (Fig. 1A): the antipolar $P4_2/mnm$ structure appears as the intermediate phase. This temperature-induced polar-antipolar transition has never been realized in any oxide HIF multiferroics/ferroelectrics. Given such transition involves the loss of the $c^+$ rotation (which does not affect ferromagnetism) and a change in the $X_3^-$ tilt direction from $a^-a^-$ ($X_3^-$ (a;0)) to $a^-b^0$ ($X_3^-$ (a;a)), we anticipate control of ferromagnetism (Supplementary Information Fig. S28). However, the polar-antipolar transition temperature ($T_C$) for $K_3Mn_2Cl_7$ is higher than the magnetic ordering temperature $T_N$, hence dipolar control of the magnetism cannot be directly realized in this composition. Substituting $K^+$ with the larger $Rb^+$ cation reduces the size-mismatch with manganese and stabilizes the antipolar $P4_2/mnm$ structure, lowering $T_C$ while keeping $T_N$ almost unchanged (Supplementary Information Figs. S22, S29). Consequently, $T_C$ can be chemically tuned below $T_N$ by varying Rb concentration. Large concentrations even stabilize the antipolar $P4_2/mnm$ as the ground-state structure as supported by our NPD data, e.g., $K_2RbMn_2Cl_7$ adopts an antipolar $P4_2/mnm$ structure at 5 K (Supplementary Information Figs. S12-S14). The 5 K NPD data also reveals a similar G-type antiferromagnetic order along the $c$-axis in magnetic space group $Pnn'm'$ in $K_2RbMn_2Cl_7$. The $a^-b^0$ tilt ($X_3^-$ irrep) creates additional canted moments in the $ab$ plane due to $DM$ interactions, evidenced by non-linear, hysteretic magnetization-isotherms (Supplementary Information Figs. S15, S16). For lower Rb concentrations, such as $K_{2.85}Rb_{0.15}Mn_2Cl_7$, we find upon cooling below $T_N$ (~ 65 K) that a structural transition from the antipolar $P4_2/mnm$ to polar $A2_1am$ occurs, accompanied by a spin-structure transition from $A2_1'am'$ to $Pnn'm'$ (Fig. 4A), evidenced by SXRD, magnetic and NPD data (Fig. 4B, Supplementary Information Figs. S19-S25). During this process both electric polarization and weak-ferromagnetic moments are reorientated in quasi-2D layers. Therefore, thermally induced simultaneous manipulation of ferromagnetism and polarization through a change in the octahedral tilts is achieved in a single phase (Supplementary Information Figs. S26-S28). Such tilt-enabled magnetoelectric coupling has long been sought but never realized. We expect a similar effect over a wide composition range in $K_{3-x}Rb_xMn_2Cl_7$ solid-solutions when the polar-antipolar transition occurs below $T_N$ (Fig. 4C).

For decades, research in magnetoelectric multiferroics has focused on using electric or magnetic field to control ferroic orders. We have shown that the concept can be thermally accessible. In $K_{3-x}Rb_xMn_2Cl_7$, simultaneous reorientations (but not switching) of electric and ferromagnetic polarizations are achieved directly using thermal energy without thermomechanical mediation. Such temperature-induced manipulation can be easily achieved remotely, which is appealing for *in vivo* application such as drug deliveries[27], or even by using 'side effects' such as intentional Joule heating. In addition, inducing a temperature difference is generally easier than applying a large electric or magnetic field which is normally required for polarization switching in current multiferroics. Since structural transitions are sensitive to many external conditions such as pressure[28] and radiation[29], one would naturally expect these stimuli to be able to trigger magnetoelectric effects.

This work presents the first example of correlated HIF multiferroics designed from halides and offers a strategy to realize new multifunctional materials in other fluorides and bromides with transition metals beyond manganese. We showed cooperative phenomena can emerge in non-oxide compounds, and there are other 'hidden' properties to be revealed. Further efforts should be devoted to discover materials hosting polar-antipolar structural transitions with larger ferromagnetic moments and higher $T_N$, with layered oxyhalides as a promising direction. It would be interesting to simultaneously apply electric fields and thermal changes around $T_C$, ideally on high-quality single-crystalline samples to check the possibility of reaching more than two multiferroic states in a single-phase material. The uniaxial NTE behaviour along the polar axis also suggests another avenue toward new NTE materials. There have been long interests in introducing electric polarization into 2D magnetic materials[30], and our results suggest that this may be achieved by incorporating alternative or multiple anions to induce a polar distortion.

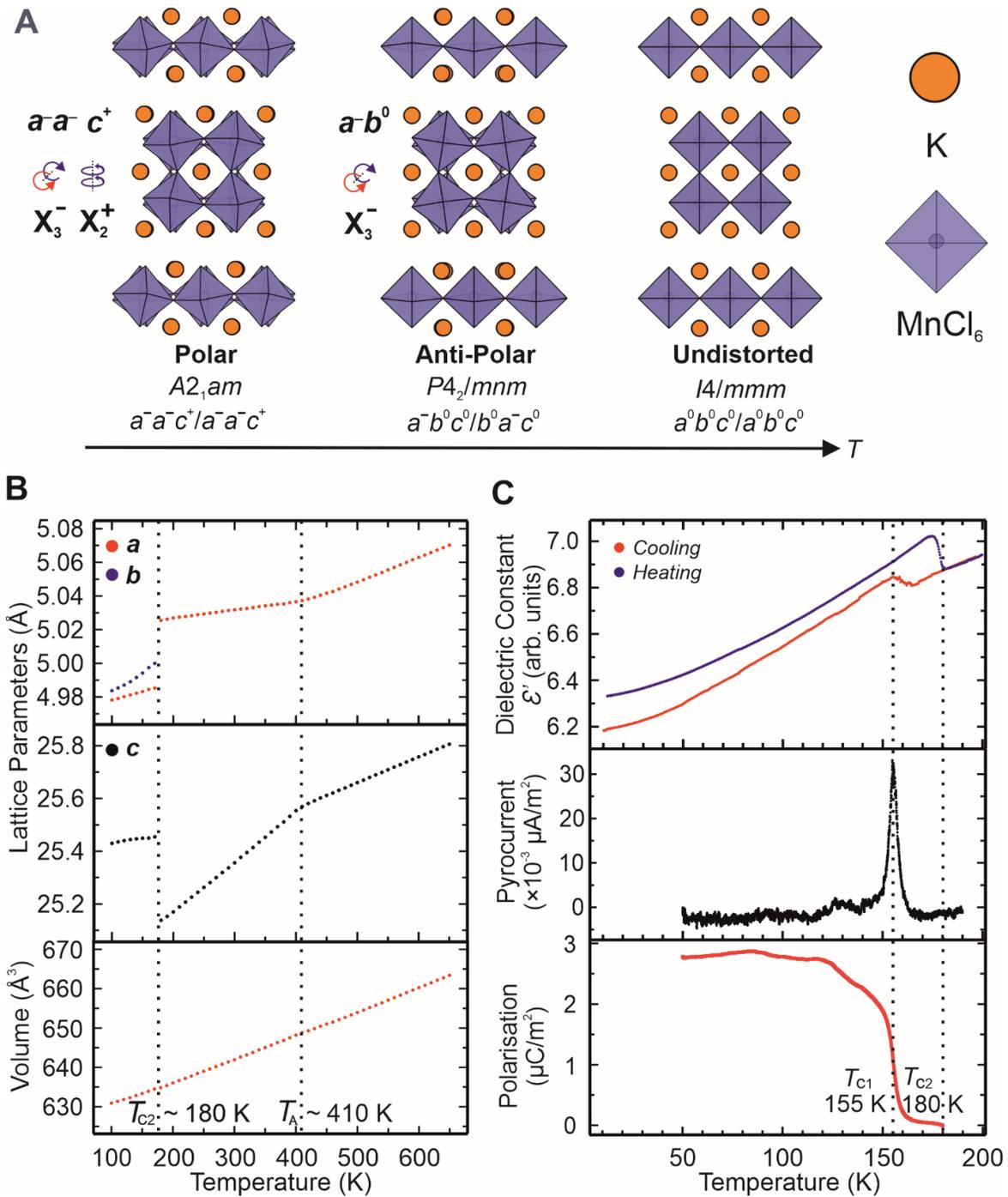

**Fig. 1. Structure evolution of K₃Mn₂Cl₇.** (**A**) Schematic of tilting distortions adopted by K₃Mn₂Cl₇ with temperature. (**B**) Normalized lattice parameters with temperature. (**C**) Temperature-dependent dielectric constants, pyroelectric currents and electric polarizations on a polycrystalline pellet.

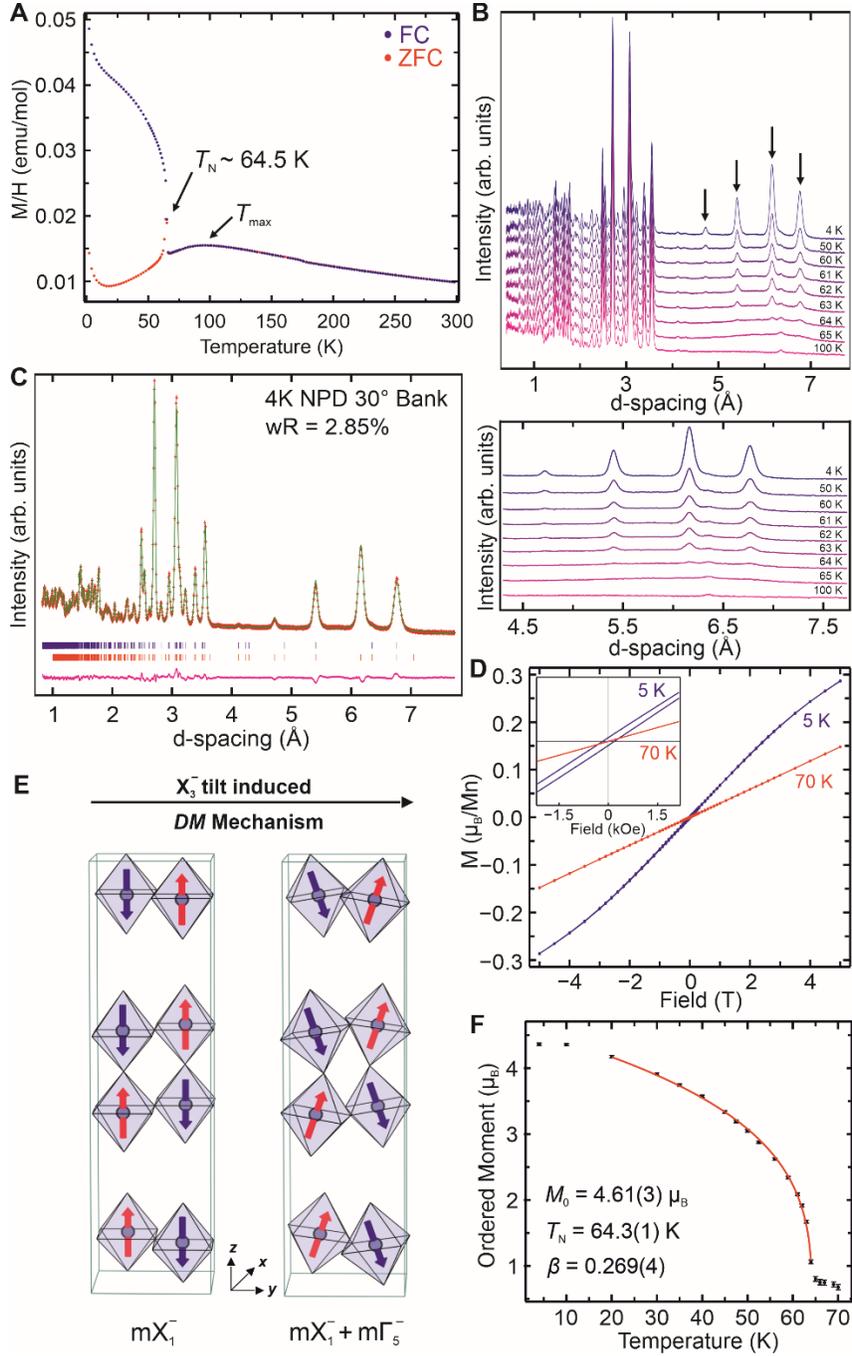

**Fig. 2. Magnetic properties of $K_3Mn_2Cl_7$.** (**A**) ZFC/FC data collected between 2 K and 300 K. (**B**) NPD data collected between 4 K and 100 K. Arrows indicate selected magnetic reflections. (**C**) Refinement of $A2_1am$ crystal and $A2_1'am'$ magnetic models against 4 K NPD data. (**D**) $M$(H) data collected at 5 K (blue) and 70 K (red). (**E**) G-type antiferromagnetic order observed from NPD data (left) and final canted-antiferromagnetic structure (right) due to *DM* interaction. (**F**) Power law fitting to ordered Mn moments against temperature.

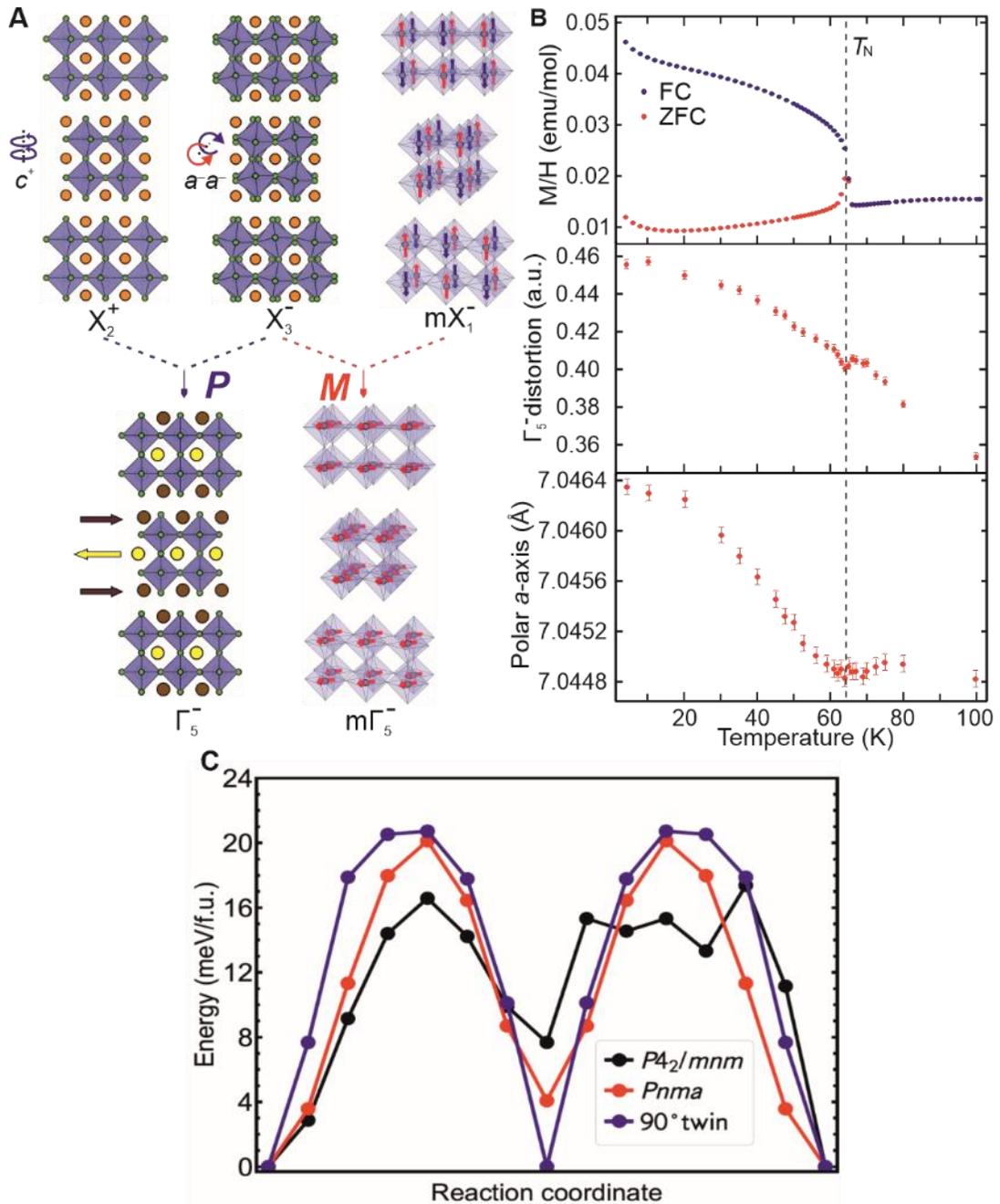

**Fig. 3. Magnetoelectric coupling in K$_3$Mn$_2$Cl$_7$.** (**A**) Magnetoelectric (ME) coupling mechanism in K$_3$Mn$_2$Cl$_7$. Switching the X$_3^-$ distortion switches both the electric polarization $P$ ($\Gamma_5^-$) and weak ferromagnetism (m$\Gamma_5^-$). (**B**) Evidence for ME coupling. Top, middle, and bottom figures show magnetization, refined polar $\Gamma_5^-$ distortion and polar $a$-axis against temperature. (**C**) Polarization energy switching barriers against reaction coordinate (i.e., changing from +$P$ state (very left point) to -$P$ state (very right point) through the intermediate phase) for different paths.

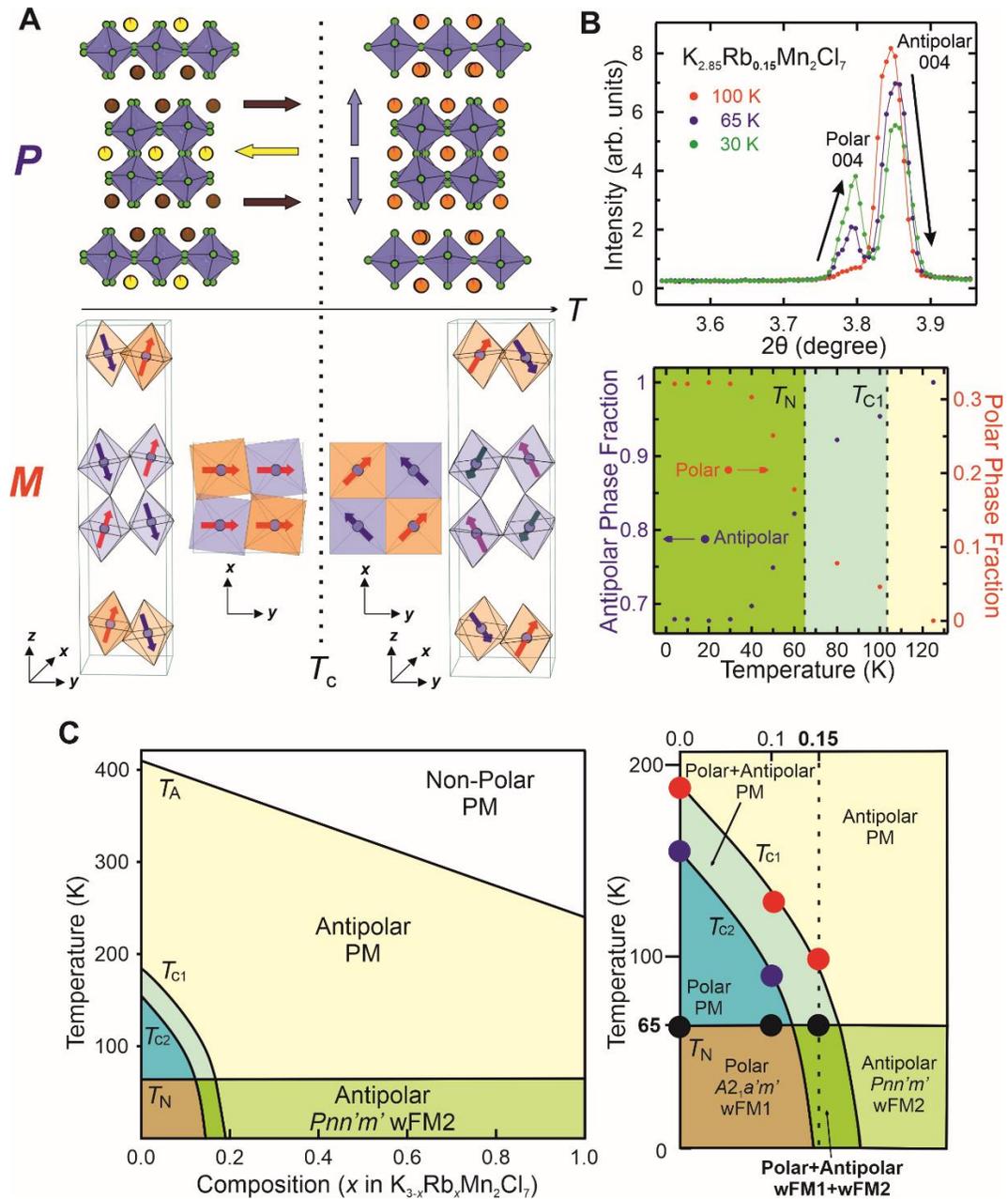

**Fig. 4. Simultaneous thermal manipulation of *P* and *M* in K$_{3-x}$Rb$_x$Mn$_2$Cl$_7$.** (**A**) Illustration of thermal changes in the directions of *P* and *M*. Arrows in the top panel indicate the local electric polarizations. (**B**) Temperature evolution of 004 reflections (SXRD) and phase fractions from $A2_1am$ and $P4_2/mnm$ phases of K$_{2.85}$Rb$_{0.15}$Mn$_2$Cl$_7$. Upon cooling, there is a transition from the antipolar to polar state. (**C**) A proposed, hypothetical phase diagram as a function of composition and temperature. The right-hand-side figure shows an expanded region around the origin. The dashed line indicates K$_{2.85}$Rb$_{0.15}$Mn$_2$Cl$_7$.

## Methods

**Synthesis**

Polycrystalline samples of $K_{3-x}Rb_xMn_2Cl_7$ ($0 \leq x \leq 1$) were prepared by reacting suitable stoichiometric ratios of anhydrous KCl (99.0%), RbCl (99.8%) and $MnCl_2$ (99.999%). The mixture was ground in an agate pestle and mortar in a nitrogen-filled glovebox, pelletized and loaded in a Pyrex tube. The Pyrex tube was then evacuated, sealed and heated to 410 °C at a rate of 100 °C/h and kept at 410 °C for 48 h. Samples were then transferred into the glovebox, reground, pelletized and heated in an evacuated Pyrex tube at 410 °C for another 48 h. Sample purity and reaction progress were monitored by X-ray powder diffraction data, which were collected using a Rigaku SmartLab SE diffractometer (Cu Kα radiation). Samples were kept in the glovebox to avoid exposure to air or moisture.

**Structural Characterization**

High-resolution synchrotron X-ray powder diffraction data were collected using instrument BL02B2 at SPring-8. Samples were packed and sealed in 0.5 mm diameter borosilicate glass capillaries. Diffraction patterns were measured using $CeO_2$-calibrated X-ray radiations with an approximate wavelength of 0.42 Å. Diffraction data were collected as a function of temperatures using a $N_2$ gas flow device ($100 \leq T/K \leq 500$) and Helium gas flow temperature control device ($30 \leq T/K \leq 100$). Variable-temperature neutron powder diffraction data were collected using SPICA instrument (Time-Of-Flight, TOF) at J-Parc facility and HERMES diffractometer (Constant Wavelength, CW) at Japan Research Reactor-3. Rietveld profile refinements were performed using the GSAS[31], GSAS2[32] and FullProf[33] suites of programs. Symmetry and distortion mode analysis were performed using the ISODISTORT software[34,35].

**Magnetic Measurements**

Magnetization data were collected using a Quantum Design MPMS-XL SQUID magnetometry. Zero-field cooled (ZFC) and field cooled (FC) data were collected in an applied field of 100 Oe. Magnetization-field isotherms were collected with applied fields in the range $-5 \leq H/\text{T} \leq 5$.

**Dielectric and pyroelectric Measurements**

For the measurements of dielectric constant ($\varepsilon'$) and electric polarization ($P$), a sintered polycrystalline sample was pressed into thin plates and, subsequently, a gold electrode was deposited by a sputtering method on a pair of the widest surfaces. The $\varepsilon'$ was measured at an excitation frequency of 100 kHz in a commercial $^4$He cryostat using an LCR meter (Agilent E4980). Electric polarization was obtained by integrating a pyroelectric current measured with an electrometer (Keithley 6517).

**Optical Second Harmonic Generation (SHG) Measurements**

For the purpose of investigating the noncentrosymmetric nature of $K_3Mn_2O_7$, optical SHG was measured at 10 K in reflection geometry using a 1064-nm fundamental beam of a Nd:YAG laser (EKSPLA PL2143) operating with 25 ps pulses at a repetition frequency of 10 Hz. The SHG light from the polycrystalline sample was detected with a photomultiplier tube through a 532-nm narrow band-pass filter. The sample was mounted in a closed-cycle helium refrigerator.

**Computational Details**

Our total energy calculations are based on density functional theory (DFT) within the generalized gradient approximation (GGA) utilizing the revised Perdew-Becke-Erzenhof functional for solids (PBEsol)[36] implemented in the Vienna Ab Initio Simulation Package (VASP)[37,38]. We use a 550-eV plane wave cutoff energy for all calculations and the projector augmented wave (PAW) method[39,40]. We also use a 5×5×2 *k*-point mesh and Gaussian smearing (0.10 eV width) for the Brillouin-zone integrations. The electric polarization is computed based on linear response theory using Born effective charges and small ionic displacements with respect to a centrosymmetric reference structure[41], which is consistent with that computed by the Berry phase method[42,43]. To obtain converged results, the planewave cutoff energy is increased to 700 eV in the calculations of Born effective charges. The DFT plus Hubbard *U* method[44] is used with the Hubbard *U* and the exchange parameter *J* set to 6 and 1 eV for Mn, respectively, which reproduced the experimentally G-type antiferromagnetic structure. Spin-orbital coupling (SOC) is included in the magnetic anisotropy studies and the *J* value is increased to 1.5 eV to reproduce the experimentally determined magnetic anisotropy.

To compute the symmetric spin exchange parameters, the four-state mapping method is used[45,46]. In this study, we calculate the effective symmetric spin exchange parameters, which are obtained by setting $|S_i|=1$, namely, $J_{ij}=J_{ij}^{eff}S_iS_j$ for a spin dimer *ij*.

Our parallel tempering Monte Carlo (PTMC) simulations are based on an exchange MC method[47], which can simulate the classical Heisenberg spin system with a Hamiltonian of $E = E_0 + \sum_{\langle i,j \rangle} J_{ij} \mathbf{S}_i \cdot \mathbf{S}_j$ where $J_{ij}$ is the symmetric spin exchange parameter. To obtain the plot of the specific heat (*C*) versus temperature (*T*), we calculate the specific heat $C \sim (\langle E^2 \rangle - \langle E \rangle^2)/T^2$ after the system reaches equilibrium at a given temperature *T* in the simulation. Then the critical temperature can be obtained by locating the peak position in the $C(T)$ plot. In our PTMC simulations of the effective Hamiltonian, a 6×6×3 supercell of the 48-atom unit cell is adopted for the $K_3Mn_2Cl_7$ compound, which is converged and only leads to changes of 2 K compared to a larger 12×12×2 supercell. The number of replicas is set to 96.

**Acknowledgments:** The authors thank A. Kaminaka from Tohoku University for sample preparation, S. Kawaguchi and S. Kobayashi from SPring-8, JASRI for the assistance in collecting SXRD data (2020A0822). NPD experiments were performed at J-Parc BL09 SPICA beamline (2019S10) and JRR-3 HERMES beamline (22602). H. K. acknowledges JSPS Core-to-Core Program (JPJSCCA202200004) and JSPS Grants-in-Aid for Scientific Research (22H04914, 22H0514). T. Z. and S. K. acknowledges GIMRT Program of the Institute for Materials Research,

Tohoku University (202109-RDKGE-0104). X. Z. and J. M. R. were supported by the National Science Foundation under award number DMR-2011208. Computational efforts were supported by the National Energy Research Scientific Computing Center (NERSC), a U.S. DOE Office of Science User Facility located at Lawrence Berkeley National Laboratory, operated under Contract No. DE-AC02-05CH11231, and the Extreme Science and Engineering Discovery Environment (XSEDE), which is supported by NSF (ACI-1548562). T. A. was supported by the JSPS KAKENHI (20K14396). K. F. was supported by the JSPS KAKENHI (JP20K20546 and JP22H01775), the Kazuchika Okura Memorial Foundation, the Nippon Sheet Glass Foundation for Materials Science and Engineering, and the Mitsubishi Foundation.

**Author contributions:** T. Z., X. Z., J. M. R., and H. K. designed the project. T. Z., S. K. and A. Y. synthesized samples. X. L. performed theoretical calculations. T. Z., Y. N., T. S. and C. T. collected SXRD and NPD data. T. Z. performed structural analysis, with comments from Y. N., K. F. and X. L. T. A. and K. O. performed physical property measurements. K. F., T. K. and T. T. performed SHG measurements. T. Z., H. T. and H. L. collected and analyzed magnetization data. T. Z., X. L., J. M. R., and H. K. wrote the manuscript, with comments from other authors.

**Competing interests:** The authors declare that they have no competing interests.

**Supplementary information** is available for this paper.

**Correspondence and requests for materials** should be addressed to James M. Rondinelli and Hiroshi Kageyama.